\documentclass[aps,prd]{revtex4b3}
\usepackage{epsfig}
\usepackage{latexsym}
\usepackage{multicol}

\newcommand{\be}{\begin{equation}}
\newcommand{\ee}{\end{equation}}
\newcommand{\bea}{\begin{eqnarray}}
\newcommand{\eea}{\end{eqnarray}}

\def\vk{{\bf k}}
\def\vr{{\bf r}}

\def\d{\partial}
\def\e{{\rm e}}
\begin{document}
  \begin{flushright} \begin{small}
     DTP-MSU/00-06 \\
     hep-th/0006087
  \end{small} \end{flushright}
\vspace{.5cm}
\title{Gravitational and dilaton radiation from a
relativistic membrane}
\author{Dmitri V. Gal'tsov}
\email{galtsov@grg.phys.msu.su}
\affiliation{ Moscow State University, 119899, Moscow, Russia}
\author{ Elena Yu. Melkumova }
\affiliation{ Moscow State University, 119899, Moscow, Russia}
\date{\today}

\begin{abstract}
Recent scenarios of the TeV-scale brane cosmology suggest a possibility
of existence in the early universe of two-dimensional topological defects:
relativistic membranes. Like cosmic strings, oscillating membranes could emit
gravitational  radiation contributing to a stochastic background
of gravitational waves. We calculate dilaton and gravitational radiation from
a closed toroidal membrane excited along one homology cycle. The
spectral-angular distributions of dilaton and gravitational radiation are
obtained in a closed form in terms of Bessel's functions. The angular
distributions are affected by oscillating factors due to an interference of
radiation from different segments of the membrane. The dilaton radiation power
is dominated by a few lower harmonics of the basic frequency, while the
spectrum of the gravitational radiation contains also a substantial
contribution from higher harmonics. The radiative lifetime of the membrane
is determined by its tension and depends weakly on the ratio of two radii of
the torus. Qualitatively it is equal to the ratio of the membrane area at
the maximal extension to the gravitational radius of the membrane as a whole.
\end{abstract}
\pacs{04.20.Jb, 04.50.+h, 46.70.Hg} \maketitle

\begin{multicols}{2}

\section{Introduction}

Now it is well understood that topological defects which could have been
created at the early stages of cosmological expansion may leave their
footprints in the spectrum of the background gravitational waves
\cite{ViSh92,BaCaSh97} which can be detected in the future experiments.
Until recently main attention was paid to the effects due to cosmic
strings. Cosmic strings, presumably serving as seeds for the formation of
galaxies, were thought to be produced at the phase transition of the grand
unification scale \cite{Ki76,HiKi95}. Gravitational, dilaton and axion
radiation emitted by oscillating strings was thoroughly studied in a number
of papers (see e.g. \cite{Vi85,Bu85,VaVi85,GaVa87,DaVi97} and references
therein). Although a recent analysis of the available astrophysical data
seems to rule out the cosmic string scenario of the formation of structures
in the Universe, these calculations  may still be of interest in
the context of the fundamental string/M theory.

Recently it was argued that the mass scale of the superstring/M theory may be
as low as 10 TeV \cite{ArDiDv98,AnArDiDv98} at the expense
of existence of some non-small extra dimensions. If true, this may drastically
modify the full scenario of cosmological evolution. Based on this
assumption various cosmological models were envisaged in which the observed
Universe is treated as a three-brane embedded in a higher-dimensional world.
According to this brane cosmology the standard model
particles are confined to the brane while gravity lives in the full space.
Such scenarios appeal to reconsider the problem of topological defects
which should be associated rather with solitons of the fundamental string
theory than with those of the field-theoretical models of grand unification.

Among the string theory solitons which may live on a three-brane
there are one-dimensional strings and two-dimensional {\it membranes}.
In the zero-thickness limit the dynamics of solitonic strings is
governed by the Nambu-Goto action, while the membranes are described by a
similar three-dimensional geometrical action (see e.g. \cite{CoTu76,Su83}).
Relativistic membrane in the four-dimensional spacetime geometrically are
similar to domain walls encountered in the usual field-theoretical models.
However their formation and dynamics in the string/brane cosmology are likely
to be different from dynamics of the domain walls in the standard
cosmological scenario. Here we do not discuss details of creation and
evolution of membranes in the brane cosmology
but concentrate on the problem of radiation.

Radiation produced by  one-dimensional defects -- the relativistic strings --
is likely to be similar to radiation from the cosmic strings, apart from a
different attribution of scalar and two-form fields involved.
Radiation from an oscillating two-dimensional relativistic membrane seems
not to have been studied in detail before, although some earlier
estimates of gravitational radiation emitted by the cosmic
domain walls are worth to be mentioned \cite{GlRo98,KoTuWa92}. Massless
boson modes to which the membranes are coupled in the four-dimensional
space-time include a dilaton, a graviton and a Ramond--Ramond three-form.
The latter is non-dynamical in four dimensions, so oscillating membranes
may emit dilatons and gravitons.

In the classical limit one can describe radiation from a vibrating membrane
in a standard way using the retarded solutions of the corresponding wave
equations. Unfortunately, contrary to the case of strings, a general
solution of the (highly non-linear) membrane equation of motion can hardly be
obtained in a closed form. To simplify the problem one can restrict oneself
by some particular excitation mode. Here we study the dilaton and
gravitational radiation emitted by a toroidal oscillating membrane excited
along a homology cycle with the radius much smaller than the second
radius of the torus. Under such an assumption the membrane equation of motion
can be easily solved, and a closed analytic expression for the spectral
power of radiation can be obtained. We analyse in detail the spectral-angular
distributions with an emphasis on the effects due to two-dimensional
nature of a membrane. Hopefully this simple model correctly describes basic
features of the membrane radiation in more general case,
although radiation due to more complicated excitation modes
is certainly worth to be considered in the future.

The plan of the paper is as follows. In Sec. II we derive the linearized
equations for the scalar and gravitational radiation from a membrane source.
In Sec. III the solution of the equations of motion for a
free toroidal membrane is obtained under an assumption of the relative
smallness of the radius of an excited homology cycle. Sec. IV is devoted to
an analysis of the dilaton radiation; we derive the spectral-angular
distribution in terms of
the Bessel's functions and investigate the role of higher harmonics of
the basic oscillation frequency. A simplified expression valid in the
high-frequency range of the spectrum is obtained in terms of the MacDonald's
functions. In Sec. V the gravitational radiation is studied along similar
lines. The results are summarized and discussed in Sec. VI.
Some useful integrals involving Bessel's functions
used in the calculation are given in the Appendix.

\section{Basic equations}
Consider a relativistic membrane
$x^{\mu}=z^{\mu}(\sigma^a),\;\mu=0,1,2,3,$ propagating in the
four-dimensional space-time with the metric $g_{\mu\nu}$ and interacting
with the dilaton  field $\phi$ (our signature convention is $(+---)$).
The corresponding action can be written as a sum
\be\label{fac}
S=S_m+S_f
\ee
of the purely geometrical membrane action \cite{CoTu76,Su83}
\be \label{mac}
S_m=\mu\int\limits \e^{-\alpha\phi}\sqrt{\gamma}\,d^3\sigma,
\ee
and the field term
\be\label{fgac}
S_f=\frac{1}{16\pi G}\int\limits\left\{-R+2
\d_\mu\phi \d_\nu\phi g^{\mu\nu}\right\}\sqrt{-g}d^4x.
\ee
Here $\mu$ is the membrane tension parameter of a dimension
of the energy per unit area, $\alpha$ is the  dilaton coupling constant.
In our normalization the dilaton field $\phi$ and the coupling constant
$\alpha$ both are dimensionless. The membrane world-volume is
parameterized by the internal coordinates
$\sigma^a =(\tau, \sigma, \rho),\quad a=0,1,2$. The induced metric on the
world-volume reads
\be \label{im}
\gamma_{ab}=g_{\mu\nu}\frac{\partial z^\mu}{\partial{\sigma^a}}
\frac{\partial z^\nu}{\partial{\sigma^b}},
\ee
where $\gamma=\det \gamma_{ab}$.

Variation of the action (\ref{mac}) over the space-time
coordinates $z^\mu$ gives the equation of motion
\be\label{meq}
\frac{1}{\sqrt{\gamma}}\d_a\left\{\e^{-\alpha\phi}\sqrt{\gamma}\gamma^{ab}
\d_b\left(z^\nu g_{\mu\nu}\right)\right\}=0,
\ee
where $\gamma^{ab}$ is the inverse metric on the world-volume. Up to the
dilaton factor, this is a three-dimensional covariant D'Alembert equation.
It is worth comparing it with the two-dimensional string equation of motion.
In the latter case one has three gauge degrees of freedom corresponding
to two diffeomorphisms and a Weyl transformation on the world-sheet.
Consequently, three independent components of the world-sheet metric can
be chosen flat, so that the equation of motion becomes linear. A general
solution can be written as a sum of two arbitrary functions of null
coordinates on the world-sheet: right and left movers. This is no more
possible in the case of a membrane for which the number of gauge degrees of
freedom is again three (diffeomorphisms of the world-volume), while the
number of the independent world-volume metric components is six. Therefore
the membrane equation of motion remains essentially non-linear in any gauge.
To make it explicitly solvable one needs additional assumptions to be
made about excitation modes.

Variation of the full action (\ref{fac}) with respect to the dilaton $\phi$
gives the four-dimensional D'Alembert equation
\be\label{deq}
\frac{1}{\sqrt{-g}}\d_\mu\left(\sqrt{-g}g^{\mu\nu}\d_\nu\phi\right)
=4\pi j(x)
\ee
with the source term
\be\label{ds}
j(x)=G\mu\alpha \e^{-\alpha\phi} \int\limits
\frac{\delta^4\left(x-z(\sigma)\right)}{\sqrt{-g}} \sqrt{\gamma}\,d^3\sigma.
\ee
Note that the effective coupling of the dilaton to the membrane is
proportional to the dilaton exponential. In absence of the dilaton
background this factor is of course negligible in the linearized limit
relevant to the radiation problem.

Finally, varying the action (\ref{fac}) with respect to
the space-time metric $g_{\mu\nu}$, one obtains
the Einstein equations
\be\label{eeq}
R_{\mu\nu}-\frac12 g_{\mu\nu}R=8\pi G \left( T_{\mu\nu}+T^{dil}_{\mu\nu}\right),
\ee
with the source term consisting of the membrane stress--tensor
\be\label{mst}
T^{\mu\nu}=\mu\int\limits\e^{-\alpha\phi} \gamma^{ab}
\d_a z^\mu \d_b z^\nu \frac{\delta^4\left(x-z(\sigma)\right)}
{\sqrt{-g}} \sqrt{\gamma}\,d^3\sigma,
\ee
and the dilaton term. In what follows we neglect the interaction
of the dilaton with gravity and omit the second term.
The effective coupling of the membrane  to gravity  contains again
the dilaton exponential factor which may be neglected in absence of
the dilaton background.

The action (\ref{fac}) is invariant under the reparameterization
of the world-volume $\sigma^a\to\sigma^{a'}(\sigma^b)$, so
three gauge conditions may be imposed on the world-volume
metric $\gamma_{ab}$. We will treat the gravitational
field perturbatively on the flat background
\be\label{met}
g_{\mu\nu}=\eta_{\mu\nu}+ h_{\mu\nu},
\ee
where $\eta_{\mu\nu}$ is the Minkowski metric, and choose the parameterization
of the world-volume ensuring diagonality of the metric $\gamma_{ab}$:
\begin{eqnarray}
 \dot{z}^\mu z'_\mu  =  0, \quad {\dot z}^\mu \bar{z}_\mu  =  0,
 \quad z'^\mu\bar{z}_\mu  = 0.
 \end{eqnarray}
Here and below all the contractions are performed using the flat
metric $\eta_{\mu\nu}$, and the following abbreviations are used
$ \dot{z}^\mu  =  \partial_\tau z^\mu, \;
 z'^\mu = \partial_\sigma z^\mu, \ \bar{z}^\mu  =
  \partial_\rho z^\mu $.
The induced metric then takes the form
\begin{eqnarray}
{\gamma_{ab}} =
\left( \begin{array}{ccc}
\dot{z}^{2} & 0 & 0 \\
0 & z'^{2} & 0\\
0 & 0 &  \bar{z}^{2}
\end{array} \right),
\quad
\gamma= \sqrt{\dot{z}^{2}z'^{2} \bar{z}^{2}},
\end{eqnarray}
where  $\dot{z}^{2}=\dot{z}^{\mu}\dot{z}_{\mu}$ etc. Note
that $\dot{z}^\mu$ is a timelike vector, while $z'^\mu$ and
$\bar{z}^\mu$ are spacelike, so the signature of $\gamma_{ab}$ is
$+--$ (hence $\det\gamma_{ab}$ is positive).
In the diagonal gauge the flat-space membrane equation of motion (\ref{meq})
can be presented as the continuity equation (see, e.g.,
\cite{CoTu76}):
\be \label{eq}
\frac{\partial P_\tau^\mu}{\partial\tau}+
\frac{\partial P_\sigma^\mu}{\partial\sigma}+
\frac{\partial P_\rho^\mu}{\partial\rho}=0,
\ee
where $P^\mu_a$ are generalized momenta obtained by varying
the membrane action over $\d x_\mu/\d \sigma^a$:
\bea \label{eqp}
P_{\tau}^\mu & = &\frac{1}{\sqrt{\gamma}} \dot{z}^{\mu}\mu z'^{2} \bar{z}^{2},\nonumber\\
P_{\sigma}^\mu & =& \frac{1}{\sqrt{\gamma}} z'^{\mu} \dot{z}^{2} \bar{z}^{2}, \\
P_{\rho}^\mu &= &\frac{1}{\sqrt{\gamma}} \bar{z}^{\mu} \dot{z}^{2} z'^{2}\nonumber.
\eea
These equations are still highly non-linear.
They can be solved, however, if one assumes certain symmetric configurations
of the membrane.

In absence of the gravitational background field (\ref{met}),
the dilaton equation (\ref{deq}) can
be considered in the Minkowski space-time. Denoting the flat
wave operator by a box we have
\be\label{deqf}
\Box \phi=4\pi j(x),
\ee
where the source term is
\be\label{dsf}
j(x)=\mu\alpha G\int\limits \sqrt{\dot{z}^{2}z'^{2} \bar{z}^2}
\delta^4\left(x-z(\sigma)\right) d^3\sigma.
\ee
Finally, the linearized Einstein equations for the metric
perturbations can be written in the Fock-de Donder gauge
\be\label{eeql}
\Box\psi^{\mu\nu}=16\pi G T^{\mu\nu},
\ee
where as usual
\be
\psi_{\mu\nu}=h_{\mu\nu}-
\frac12 h_\alpha^\alpha\eta_{\mu\nu},\quad \partial_\mu \psi^{\mu\nu}=0.
\ee
In the linearized theory the membrane stress-energy tensor simplifies to
the following expression
\bea\label{mstf}
T^{\mu\nu}=\mu\int\limits \left(\dot{z}^\mu \dot{z}^\nu z'^2 \bar{z}^2
+z'^\mu z'^\nu \dot{z}^2\bar{z}^2
+\bar{z}^\mu \bar{z}^\nu \dot{z}^2 z'^2\right) &&\nonumber\\
\times \delta^4\left(x-z(\sigma)\right)
\left(\dot{z}^{2}z'^{2} \bar{z}^2\right)^{-1/2}
d^3\sigma &&.
\eea
Note that the integral in the right hand side contains one integration
more as compared with the case of a string and two integrations more
compared with the point particle. The extended nature
of the membrane source gives rise to the interference effects in the angular
distribution of the radiation power.
\section{Toroidal oscillating membrane}
Our choice of the membrane model is motivated as follows. In view of
complexity of the general situation, we intend
to calculate radiation coming from the lowest excitation mode of the closed
membrane. First the topology of the membrane 2-dimensional surface
has to be specified. The simplest spherical membrane does not radiate in
the main (spherical) excitation mode. So we choose a toroidal membrane in
which case excitations naturally produce a non-vanishing quadrupole moment.
We also assume that only one homology cycle of the torus is excited, this
considerably simplifies the equations of motion.

Specifying $z$-direction of the coordinate system to be along the symmetry
axis of the torus we arrive at the following parameterization of the membrane
world-volume
\begin{eqnarray}\label{anz}
z^0&=&\tau,\;\; z^1=\left[R+r(\tau)\cos\rho\right]\cos\sigma,\nonumber\\
z^2&=&\left[R+r(\tau )\cos\rho \right]\sin\sigma,\;\; z^3=r(\tau)\sin\rho,
\end{eqnarray}
where $R=$const is the large radius and $r(\tau)$ is the variable small
radius. Two parameters on the world volume
\be
\sigma \in [0,2\pi],\quad \rho \in [0,2\pi]
\ee
measure the angular distance along large and small homology cycles
respectively, and $\tau$ is a time coordinate on the world-volume
equal to the global time in the chosen gauge.
This ansatz satisfies the equations of motion (\ref{eq}) only if
at any moment $\tau$ one has
\be  \label{Rr}
R^2 \gg r^2.
\ee
In this approximation the world-volume element is equal to
\be\label{ga}
\sqrt{\gamma}=R\sqrt{r^2\left(1-{\dot r}^2\right)},
\ee
substituting  this into the equations of motion (\ref{eq}), one finds that
under the assumption (\ref{Rr}) the full system reduces to a single
equation
\be
\ddot{r} r \ + \ (1 \ - \ \dot{r}^{2}) \ = \ 0.
\ee
This equation has the following solution
\be\label{r}
r=\Omega^{-1} \sin(\Omega\tau),
\ee
where $\Omega$ is a constant, and it is assumed for simplicity that $r(0)=0$.
The parameter $\Omega$ is the oscillation frequency which is equal to the
inverse radius of the smaller circle $r_0$ of the torus at the moments of its
maximal extension.

In our approximation each loop $\sigma=$const belonging to the membrane
moves independently. Actually the solution (\ref{r}) satisfies exactly the
equations of motion for an infinite cylindrical membrane.
For a toroidal membrane this solution is only the main approximation
valid if $R \gg r_0$.

Note that the function $r$ in (\ref{anz})
may take both positive and negative values. Indeed, two points on the loop
$\sigma=$const which correspond to opposite ends of the diameter
$\rho=0,\,\pi$ move to each other while $r(\tau)$ is decreased,
merge when $r=0$, and then exchange their positions while $r(\tau)$
is negative. Therefore a half-period of $r$-oscillations corresponds
to the full period of the true oscillations of the membrane.
In other words, the period of the membrane motion
is $\pi/\Omega$ and not $2\pi/\Omega$.

In our units the inverse angular frequency is equal to the small radius
of the torus $r_0=\Omega^{-1}$ at the moments of its maximal extension.
Therefore the quantity
\be
v=\Omega R =\frac{R}{r_0},
\ee
equal to the ratio of two radii of the toroidal
membrane at the maximal extension, may be regarded as the shape
parameter of the membrane.


\section{Dilaton radiation}
The computation of the scalar radiation starting with the D'Alembert
equation (\ref{deqf}) is standard and amounts
to the evaluation of the Fourier-transform of the current (\ref{dsf}).
Since the membrane motion is periodic with the period $\pi/\Omega$,
the radiation power can be presented as a sum over the even harmonics
of the frequency $\Omega$
\be
P=\sum_{n=1}^\infty P_{n}, \quad \omega_n=2n\Omega.
\ee
The angular distribution of radiation with the frequency $\omega_n$ is
given by

\be
\frac{dP_n}{d\Omega}=\frac{\omega_n^2}{2\pi G}\,\left|j(\vk,\,\omega_n)\right|^2
\ee
where $\vk$ is the wave vector, $\vk^2=\omega_n^2$, and the Fourier-
-transform of the source current is defined as follows
\be \label{jk}
j(\vk,\,\omega_n)=\frac{\Omega}{\pi}\int\limits_0^\pi d t\int\limits
\e^{i(\omega_n t-\vk \vr)} j(\vr,\,t) d^3x.
\ee
Parameterizing the wave vector $\vk$ by the spherical angles
$\theta, \varphi$:
\be
\vk=\omega_n (\sin\theta\cos\varphi,\sin\theta\sin\varphi,\cos\theta),
\ee
and, using in (\ref{jk}) explicit expressions (\ref{dsf}) and (\ref{anz}),
one obtains after integration over the space-time variables $t,\,\vr$:
\be
j(\vk,\,\omega_n)=\frac{\mu\alpha G R}{\pi\Omega}
\int\limits_0^\pi d\chi\int\limits_{-\pi}^{-\pi}d\rho
\int\limits_0^{2\pi}d\sigma\e^{2ni\psi}\sin^2\chi,
\ee
where
\be  \label{int1}
\psi=\left(\chi-\sin\rho\cos\theta\sin\chi
-v\cos(\sigma-\varphi)\sin\theta\right).
\ee
Note that the azimuthal angle $\varphi$ enters into this expression
(as well as to all other formulas below) only in the combination
$\sigma-\varphi$, and thus $\varphi$ can be eliminated by the shift of the
integration variable $\sigma\to\sigma-\varphi$. This means that all quantities
are actually
$\varphi$-independent, as it could be expected in view of the axial symmetry
of the membrane. Assuming without loss of generality  $\varphi=0$, one has
\be\label{int2}
\psi=\left(\chi-\sin\rho\cos\theta\sin\chi
-v\cos\sigma\sin\theta\right).
\ee

The integration in (\ref{int1}) can be performed in terms of the Bessel's
functions as follows.
First one integrates over $\chi$ using the integral representation
(\ref{A0}):
\bea \label{sin}
&&\int\limits_0^\pi \e^{2ni\left(\chi-z\sin\chi\right)}d\chi=
\int\limits_0^\pi \cos\left[2n\left(\chi-z\sin\chi\right)\right]d\chi
\nonumber\\
&&+i\int\limits_0^\pi \sin\left[2n\left(\chi-z\sin\chi\right)\right]d\chi
=\pi J_{2n}(2nz) +i I,
\eea
where $z=\sin\rho\cos\theta$.
It can shown that the contribution proportional to the imaginary part $I$
vanishes after integration over $\rho$.
Indeed, a constant shift of the integration variable
$\chi\to\chi+\pi/2$ leads to the following integral
\be
\int\limits_{-\pi/2}^{\pi/2} d\chi\int\limits_{-\pi}^{\pi}d\rho
\sin\left[2ni\left(\chi-z\cos\chi\right)\right].
\ee
Here the integrand changes a sign under reflection of
the domain of integration
$\chi\to -\chi,\, \rho\to -\rho$. Writing in (\ref{int1})
$\sin^2\chi=(1-\cos 2\chi)/2$ one can show that the $iI$ term
gives no contribution in presence of $\cos 2\chi$ as well.

The second integration over $\rho$ is performed using the formula
(\ref{sig1}) of the Appendix.
Finally, integrating over $\sigma$ and using the recurrent
relations (\ref{AJ}) one  obtains
\be \label{dcfin}
j(\vk,\,\omega_n)=-2\pi^2\mu\alpha G\frac{R}{\Omega} J_0(y)
\left(J_n^2(x)\tan^2\theta+{J'}_n^2(x) \right),
\ee
where $y= 2nv\sin\theta$ and $x=n\cos\theta$.

In this expression the Bessel's functions $J_n(x)$ are
typical for radiation from relativistic sources like the synchrotron radiation
from a relativistic charge. Another Bessel's function factor $J_0(y)^2$
has a different origin: it describes interference effects due to
coherence of radiation emitted by different segments of the membrane.
This factor reflects an extended nature of the membrane.

One can see that the angular distribution
of the total power diverges in both directions
along the symmetry axis $\theta =0, \pi$. For
$\sin\theta=0$, the interference factor $J_0(0)=1$, so
using an asymptotic formula (\ref{A5}) we obtain
for $n\gg 1$:
\be
\frac{dP_n(0)}{d\Omega}=\frac{32}{3}\left(\frac43\right)^{1/3}
\Gamma^{-4}\left(\frac{1}{3}\right)\, G \mu^2\alpha^2\,R^2\, \pi^3\,
n^{-2/3}.
\ee
The sum over $n$ diverges, but one can show that the solid angle
into which the high $n$ radiation is peaked decreases with growing $n$,
so that the total radiation power remains finite.

The  angular distribution of the $n$-th harmonic is
given by
\be
\frac{dP_n}{d\Omega}=8 G \pi^3(\mu \alpha n R)^2 J_0^2(y)
\left(\tan^2\theta J^2_n+{J'}_n^2\right)^2,
\ee
where the argument of the Bessel's functions $J_n$ and $J'_n$ is
$x=n\cos\theta$. This is shown at Figs. 1,2 for $n=1,2$ and $v=10$.
Angular distribution of the first harmonic has a maximum near the equatorial
plane. Oscillations  are due to the interference factor $J_0^2(y)$.
Recall that our assumption (\ref{Rr}) is equivalent to $v\gg 1$, so
for non-small $\theta$ one can replace
$J_0(y)$ by the asymptotic expression (\ref{Bas}):
\be\label{Jas}
J_0(y)\approx \sqrt{\frac{2}{2\pi v n \sin\theta}} \,
\cos\left(2\pi vn\sin\theta-\frac{\pi}{4}\right).
\ee

The distribution of the second harmonic exhibits maxima displaced from the
equatorial plane to some intermediate region. It is modulated by the
interference factor with doubled frequency.

Performing the angular integration one obtains the distribution over the
harmonic number
as shown at the Fig.3. One can see that about $88\%$ of total energy loss
coresponds to the first, and about $7\%$ to the second harmonic, while the
contribution from
higher harmonics is relatively small.

The high frequency tail may be
described using approximations for the Bessel's functions
which follow from the asymptotic formulas (\ref{A1},\ref{A2}) valid
for large $n$
\bea
J_n(na)&\approx&\frac{\sin\theta}{\pi\sqrt{3}}
K_{1/3}\left(\frac{n}{3}\sin^3\theta\right),\nonumber\\
J'_n(na)&\approx&\frac{\sin^2\theta}{\pi\sqrt{3}}
K_{2/3}\left(\frac{n}{3}\sin^3\theta\right),
\eea
where $K_{1/3}$ and $K_{2/3}$ are MacDonald's functions.
Since $K_\nu$ decrease exponentially for large
values of the argument (\ref{A}), it is clear that
the high-frequency part of the radiation is concentrated in a narrow angular
regions around $\theta =0,\, \pi$
\be
\sin\theta<n^{-1/3}.
\ee
This beaming is a typical relativistic effect. The main contribution
to the total power comes from the angular region in which the
argument of the Bessel's
function $J_0(y)$ is large
\be
y\sim 2\Omega R n^{1/3}\gg 1,
\ee
so the interference factor is rapidly oscillating. Therefore
one can further simplify the angular distribution for higher harmonics
using the asymptotic formula (\ref{Jas}) and averaging over
the interference ripples as follows
\be
{\overline {J_0^2(y)}} \approx \frac{1}{\pi y}.
\ee
As a result, the averaged over ripples angular distribution of radiation
for $n\gg 1$ will be given by the formula
\be \label{diK}
\frac{d{\overline P}_n}{d\Omega}=\frac{4 G\mu^2\alpha^2 Rn}{9\pi^2\Omega}\sin^7\theta
\left(K^2_{1/3}(z)+ K^2_{2/3}(z)\right)^2,
\ee
where
\be
z=\frac{n}{3}\sin^3\theta.
\ee
Note, that this representation remains qualitatively good even for
relatively small $n$, see Fig.2.

In view of a rapid convergence of the integrals from the MacDonald's
functions, we can integrate this averaged distribution over angles.
Passing to the argument of the MacDonald's functions as the new
integration variable and using the relation
\be
\sin^8\theta d\theta=\frac{9}{n^3} z^2 dz,
\ee
where an additional factor $\sin\theta$ comes from $d\Omega$, one obtains
for $n\gg 1$
\be
{\overline P}_n=\frac{32 I_1 G\mu^2\alpha^2 R}{\pi \Omega} \, \frac{1}{n^2},
\ee
where
\be
I_1=\int\limits\left(K^2_{1/3}+K^2_{2/3}\right)^2  dz=4.33,
\ee
Therefore the contribution from high harmonics falls off as $n^{-2}$.

One can also investigate the dependence of the radiated power over
the shape parameter $v$ (the ratio of two radii of the torus). Numerical
probes for different values of $v$
show a rather weak dependence of the total radiation power on this
shape factor (see Fig.4)
\section{Gravitational Radiation}
Gravitational radiation power can be computed along similar lines.
In addition one has to distinguish between the polarisation states.
The angular distribution of radiation at the frequency $\omega_n$ can
be presented as a sum over two circular polarisations

\be
\frac{dP_n}{d\Omega}=
\frac{G\omega_n^2}{\pi}\sum_{\pm}|T^\pm(\vk,\,\omega_n)|^2,
\ee
where
\be
T^\pm(\vk,\,\omega_n)=\varepsilon^\pm_{\mu\nu}T^{\mu\nu}(\vk,\,\omega_n),
\ee
\be
T^{\mu\nu}(\vk,\,\omega_n)
=\frac{\Omega}{\pi}\int\limits_0^\pi d t\int\limits
\e^{i(\omega_n t-\vk \vr)} T^{\mu\nu}(\vr,\,t) d^3x.
\ee
Here the standard chiral graviton projectors are used
\bea \label{kirpol}
\varepsilon^\pm_{\mu\nu}&=&\varepsilon^\pm_{\mu} \varepsilon^\pm_{\nu},
\nonumber\\
\varepsilon^\pm_{\mu}&=&\frac{1}{\sqrt{2}}(e^{(1)}_\mu \pm i e^{(2)}_\mu),
\eea
where $e^{(1,2)}_\mu$ are two linearly independent
space-like unit four-vectors orthogonal
to the wave four-vector $k^\mu=(\omega_n, \vk)$ and its spatially reflected
dual ${\bar k}^\mu=(\omega_n, -\vk)$,
\bea
e^{(1,2)}_\mu k^\mu&=&e^{(1,2)}_\mu {\bar k}^\mu=e^{(1)}_\mu e^{(2)\mu}
=0,\nonumber\\
e^{(1)\mu}e^{(1)}_\mu&=&e^{(2)\mu}e^{(2)}_\mu=-1.
\eea
Alternatively one could choose the linear polarisation states
$e^{\oplus}_{\mu\nu},\,  e^{\otimes}_{\mu\nu}$ corresponding to
real and imaginary parts of (\ref{kirpol}):
\be \label{linpol}
\varepsilon^\pm_{\mu\nu}=\frac{1}{\sqrt{2}}\left(e^{\oplus}_{\mu\nu}\pm i
e^{\otimes}_{\mu\nu}\right),
\ee
or, explicitly,
\bea \label{plane}
e^{\oplus}_{\mu\nu}&=&\frac{1}{\sqrt{2}}\left(e^{(1)}_\mu e^{(1)}_\nu-
e^{(2)}_\mu e^{(2)}_\nu\right),\nonumber\\
e^{\otimes}_{\mu\nu}&=&\frac{1}{\sqrt{2}}\left(e^{(1)}_\mu e^{(2)}_\nu+
e^{(2)}_\mu e^{(1)}_\nu\right).
\eea

In the Lorentz frame in which $e^{(1,2)}_0=0$ two unit vectors orthogonal
to $\vk$ can be chosen along $\theta$ and $\varphi$ directions:
\bea
{\bf e}^{(1)}&=&(\cos\theta\cos\varphi,\cos\theta\sin\varphi, -\sin\theta),
\nonumber\\
{\bf e}^{(2)}&=&(-\sin\varphi,\cos\varphi,0).
\eea
As in the dilaton case, the chiral projections of the Fourier-transform
of the stress tensor depend only on the difference
$\sigma-\varphi$, so without loss of generality we can set $\varphi=0$.
After some simplifications one finds
\be\label{intt1}
T^{\pm}(\vk,\,\omega_n)=
\frac{\mu R}{2 \pi\Omega}
\int\limits_0^\pi d\chi\int\limits_{-\pi}^{-\pi}d\rho\int\limits_0^{2\pi}d\sigma
\e^{2ni\psi}\Phi^{\pm},
\ee
where
\be \label{intt2}
\Phi^{\pm}=\left(\cos\sigma\cos\theta\cos\rho-\sin\theta\sin\rho\pm i
\sin\sigma\cos\rho\right)^2,
\ee
and $\psi$ is given again by the Eq.(\ref{int2}).

After the integration over $\sigma,\rho$ both chiral amplitudes
lead to the same result. Indeed, purely imaginary terms in $\Phi^{\pm}$
are proportional either to $\sin 2\sigma$ or to $\sin 2\rho$; taking
into account an explicit form of the phase factor $\psi$, one finds
that both these terms do not contribute to the integral
because of the antisymmetry
of the integrand in appropriate variables. In view of the relation
(\ref{linpol}) between chiral and linear polarisation projectors
this means that the gravitational radiation in any direction
has only one linear polarisation component $e^{\oplus}_{\mu\nu}$.
By similar reasoning one can show that the product of the
first two terms in (\ref{intt2}) also vanishes after an
integration over $\rho$,
so $\Phi^{\pm}$ can be replaced by the sum of the squares of each of
the three terms. After some rearrangements, we find the following
equivalent expression
suitable for further integration
\be\label{fi}
\Phi_1=\frac{\sin^2\theta}{4} (1-3\cos 2\rho)
+\frac{1+\cos^2\theta}{4}\cos 2\sigma
(1+\cos 2\rho).
\ee
Note that, contrary to the dilaton case, this function is independent on
$\chi$, ({\it i.e.} on time). Therefore the integration over
$\chi$ gives again the expression (\ref{sin}), and the imaginary part of
the integral will vanish
after the integration over $\rho$; this remains true for terms in
(\ref{fi}) containing $\cos 2\rho$. The subsequent integration
over $\rho$ is performed using the integral
(\ref{sig1}). The integration over $\sigma$ directly amounts to the
integral representation for the Bessel's functions (\ref{A}), and finally
one obtains
\bea \label{gramp}
T^+(\vk,\omega_n)&&=\frac{\pi^2\mu R}{2 \Omega}\Big[
(1+\cos^2\theta)J_2(y)\left({J'}_n^2-\tan^2\theta J_n^2\right)\nonumber\\
+\sin^2\theta && J_0(y)
\left(J_n^2\left(1+3/\cos^2\theta\right)-3{J'}_n^2\right)
\Big],
\eea
where the argument of the Bessel's functions of the order $n, n\pm 1$ is
$x=n\cos\theta$. Now an interference of radiation produced by different
parts of the membrane is accounted by two different Bessel's functions
$J_0(y)$ and $J_2(y)$; apparently this is related to the graviton spin.
Since $J_2(0)=0$,
the first term in (\ref{gramp}) vanishes at $\theta=0, \pi$, so does the
second term containing an explicit factor $\sin^2\theta$.
Thus, contrary to the dilaton case,
the amplitude of the gravitational radiation strictly
along the symmetry axis is zero.
Still the main part of radiation is concentrated near
this direction, see Figs.~5,6.

The maximal contribution to the total radiation power comes from
the first harmonic, but its relative contribution is smaller than
in the dilaton case (Fig.~7). The high frequency tail can be described
in terms of the MacDonald's functions.
For large $y$ corresponding to the main contribution to
the totat power one has $J_0\approx J_2$ (see (\ref{Bas}), this can be used
before averaging over the ripples.
An averaged distribution in the high frequency region is then obtained
as follows
\be \label{grK}
\frac{d{\overline P}_n}{d\Omega}=\frac{4 \mu^2 Rn}{9\pi^2\Omega}\sin^7\theta
\left(3K^2_{1/3}(z)- K^2_{2/3}(z)\right)^2.
\ee
After the angular integration one gets
\be
{\overline P}_n=\frac{8 \mu^2 R I_2}{\pi \Omega} \,\,
\frac{1}{n^2},
\ee
where
\be
I_2=\int\limits\left(3K^2_{1/3}-K^2_{2/3}\right)^2  dz=1.24.
\ee
This is similar to the previous result for the dilaton (\ref{diK}).

Numerical integration for different $v$ shows that the total radiation power
depends on the ratio of the torus radii even weaker than in the dilaton case
(Fig.~8).

\section{Concluding remarks}

Let us summarize our results. We have shown that the dilaton and gravitational
radiation from a vibrating  relativistic membrane exhibits features
typical for extended relativistic sources: the radiation power contains high
harmonics of the basic frequency beamed along the symmetry axis of the torus
and its angular distribution is
substantially modulated due to interference. The relative contribution
of the high frequency tail depends on the spin and it is more
pronounced in the gravitational case. For high $n$
the averaged over ripples radiation power falls off as
${\overline P}_n\sim 1/n^2$ both in the dilaton and gravitational cases.
The high frequency part of radiation
can be conveniently described in terms of the MacDonald's functions like
in the case of ultrarelativistic point particles.

The total power of the radiation from a toroidal membrane can be presented
in the form
\be
P=G\mu^2 A f(v),
\ee
where $A\sim R r_0$ is the membrane area at the maximal extension,
and the shape factor $f(v)$ is a function of the ratio of two
radii of the torus (actually we observed rather weak variation of
$f(v)$). On dimensional grounds it can be expected that this
formula will be valid for other excitation modes up to some
factor, which we set to unity $f(v)\sim 1$ for a qualitative
estimate. Then in terms of the total mass of the membrane $M=\mu
A$ we get
\be
P\sim \frac {G M^2}{A},
\ee
so the radiative lifetime may be estimated as
\be
\tau_{rad}=\frac{M}{P}\sim \frac{A}{GM}.
\ee
This quantity is proportional to the ratio of the area at the maximal
extension to the gravitational radius of the membrane as a whole
(an inverse speed of light factor has to be inserted
in the ordinary units).
In our  model the membrane area at the moments of a
maximal extension is $4\pi^2 R/\Omega$ (recall that $\Omega$ is
equal to an inverse radius of the small circle of the torus, and
we assumed that $R\gg 1/\Omega$), therefore the quantity
$\mu R/\Omega$ is proportional to the total mass. Thus the membrane
lifetime
\be
\tau\sim \frac{\mu R /\Omega}{P}\sim \frac{1}{G\mu}
\ee
depends only on the membrane tension.
In contrary, the lifetime the radiating string loop is proportional to
the loop size.

In the string case
the radiative power in terms of a total mass has the order
\be
P^{st}\sim \frac {G M^2}{L^2},
\ee
where $L$ is the string length. Correspondingly, the radiative lifetime
is
\be
\tau^{st}_{rad}=\frac{M}{P^{st}}\sim \frac{L^2}{GM}.
\ee
Actually our toroidal membrane has a shape of a 'thick' string, following
this similarity one can identify $L\sim R, \, A\sim R r_0$. Then it is
easy to see that the membrane lifetime is $R/r_0 \gg 1$ times shorter
than that of a string of an equal length.

It can be expected that gravitational radiation from membranes will depend
on the membrane topology. For a toroidal membrane the radiation coming
from the main excitation mode (as we have assumed here) in non-zero, while
in the case of the spherical membrane the main sprerically symmetric
excitation mode does not contribute to the gravitational radiation at all.

This work was supported in part by the RFBR grant 00-02-16306.

\section{Appendix}
Evaluating the integral in the representation of the Bessel's functions
\be \label{A0}
J_n(y)=\frac{1}{\pi}\int\limits_0^{\pi}\cos{(n\sigma-y\sin\sigma)}d\sigma
\ee
for $y=nx$ and $n\gg 1$ in the stationary phase approximation one finds
\be\label{A1}
J_n(nx)\approx \frac{\sqrt{2(1-x)}}{\pi \sqrt{3}}
K_{1/3}\left(\frac{2\sqrt{2}}{3}n (1-x)^{3/2}\right),
\ee
where $K$ is the MacDonald's function.
For the Bessel's functions one has the following reccurent relations
\be\label{AJ}
J_{n\pm 1}(nx)=\frac{1}{x}J_n\mp J'_n,
\ee
while for the MacDonald's functions with non-integer $\nu$
\be \label{AK}
zK'_\nu+\nu K_\nu=-zK_{1-\nu},
\ee
(note that $K_{-\nu}=K_{\nu}$).
Differentiating (\ref{A1}) and using (\ref{AK})
one obtains in the leading approximation for large $n$
the corresponding formula for the derivative of the Bessel's function
\be\label{A2}
J'_n(nx)\approx \frac{2(1-x)}{\pi\sqrt{3}}
K_{2/3}\left(\frac{2\sqrt{2}}{3}n (1-x)^{3/2}\right).
\ee
For small arguments the MacDonald's function behaves as
\be \label{A3}
K_\nu(z)\sim\frac{\pi}{2\sin\pi\nu}\Gamma^{-1}(1-\nu)
\left(\frac{2}{z}\right)^\nu,
\ee
while for large ones
\be \label{A}
K_\nu\sim\sqrt{\frac{\pi}{2z}} \e^{-z}.
\ee
For Bessel's functions at large $y$ one has
\be \label{Bas}
J_p (y)\approx \sqrt{\frac{2}{\pi y}} \cos\left(y-\frac{p\pi}{2}-\frac{\pi}{4}\right)
\ee

Passing in the Eqs.(\ref{A1},\ref{A2}) to the limit $x\to 1$, in which case
the asymptotic behavior (\ref{A3}) holds, one obtains for $n\gg 1$
\be \label{A4}
J_n(n)\approx 3^{-2/3}\Gamma^{-1}(2/3)\left(\frac{2}{n}\right)^{1/3},
\ee
\be  \label{A5}
{J'}_n(n)\approx 3^{-1/3}\Gamma^{-1}(1/3)\left(\frac{2}{n}\right)^{2/3}.
\ee
In the main text we have used the following integral:
\be \label{sig1}
\int\limits_0^\pi \cos 2\mu\rho
J_{2\nu}(2a\sin\rho)d\rho=\pi \cos\pi\mu J_{\nu-\mu} (a)J_{\nu+\mu} (a).
\ee
In particular, in view of the Eq.(\ref{AJ}) one has

\be \label{sig2}
\int\limits_0^\pi J_{2n}(2nx\sin\rho)\cos 2\rho d\rho=
\pi \left({J'}_n^{2}(nx)- \frac{1}{x^2}J_n^2(nx)\right).
\ee

\end{multicols}
\newpage
\begin{figure}
\begin{minipage}[t]{8cm}
\centerline{\epsfig{file=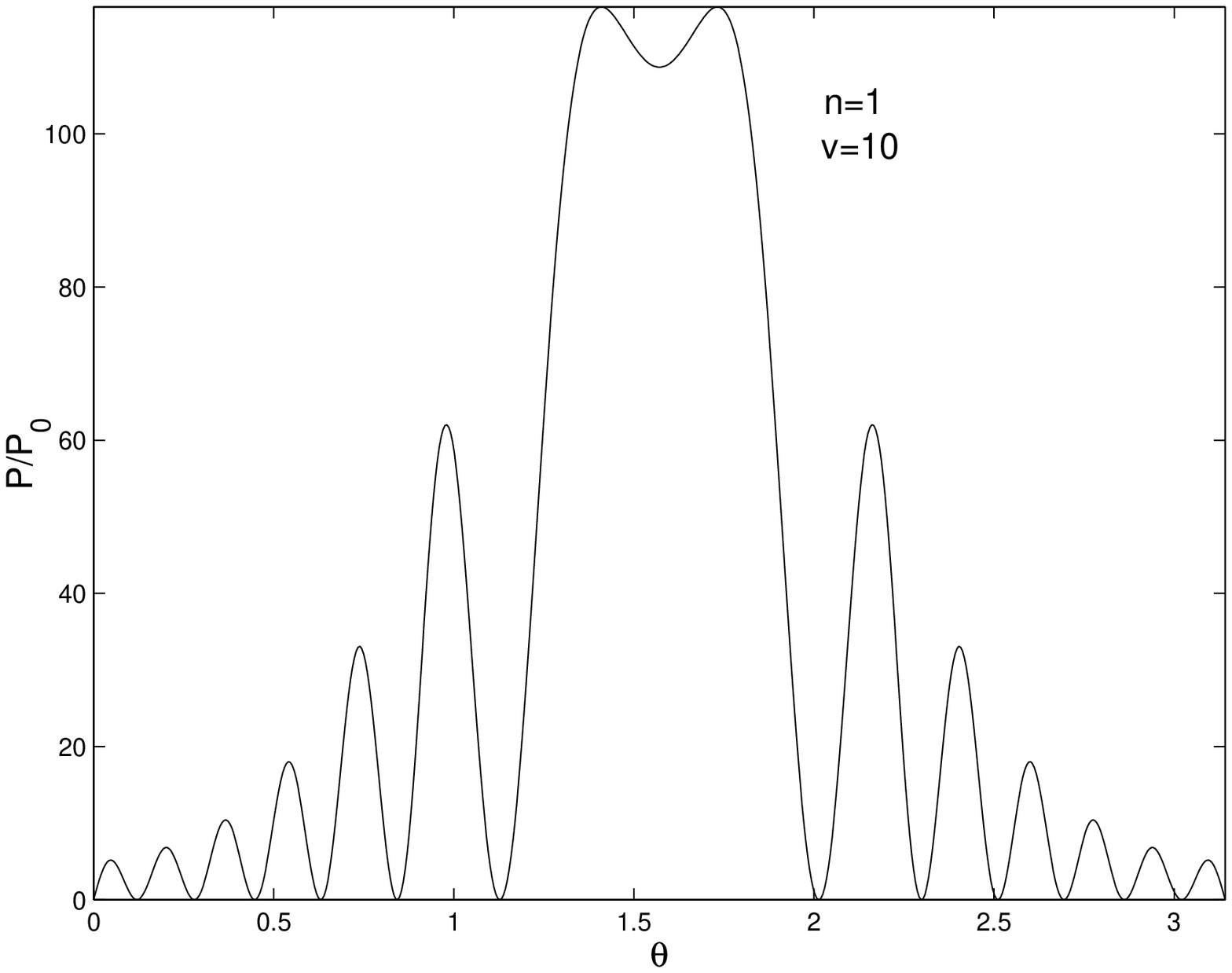,width=7cm}}
\vspace{2.5mm} \caption{Angular distribution of the first harmonic
of the dilaton radiation.}
\end{minipage}\hspace{1cm}
\begin{minipage}[t]{8cm}
\centerline{\epsfig{file=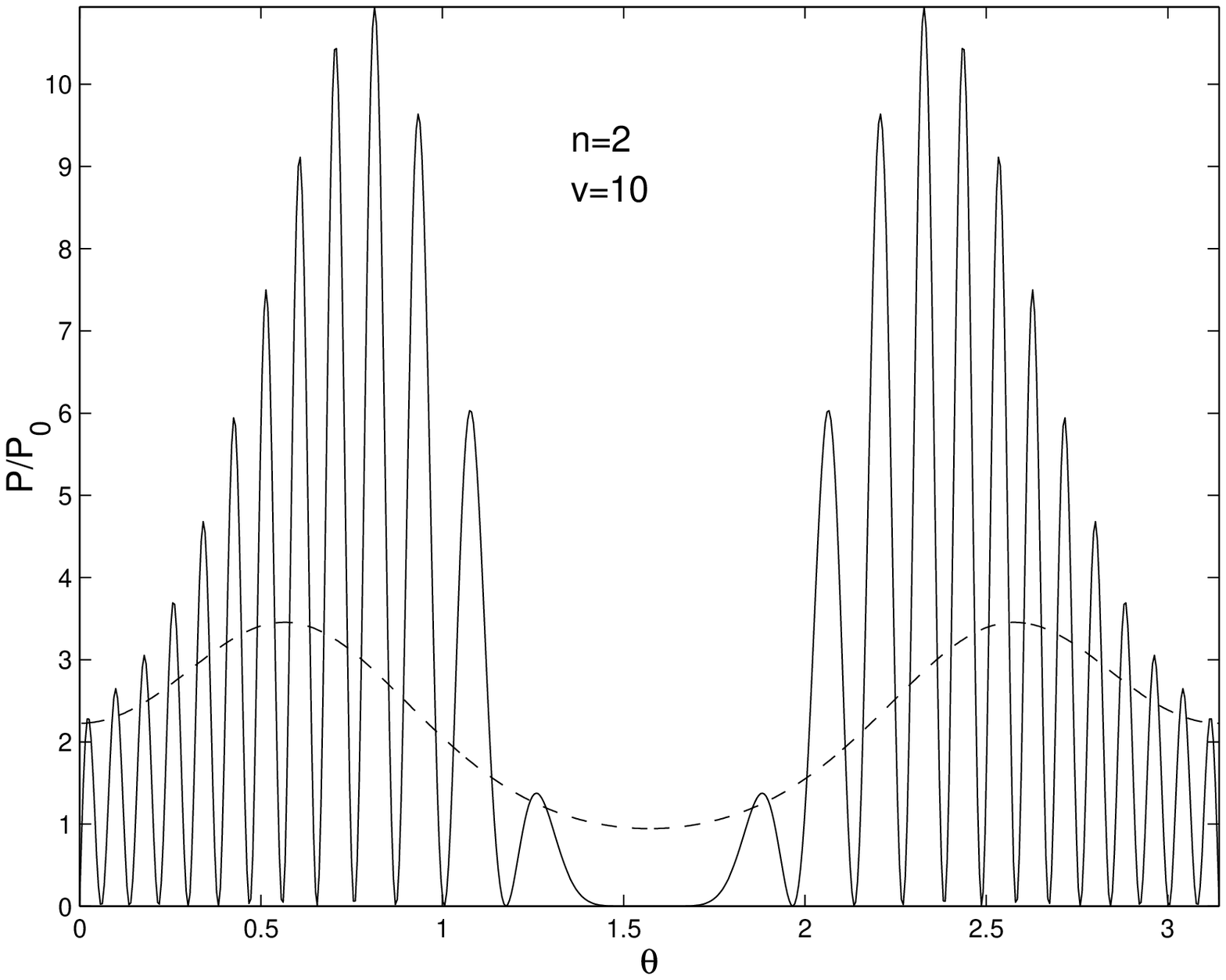,width=7cm}} \vspace{2.5mm}
\caption{Second harmonic of the dilaton radiation and its averaged
over ripples in terms of the MacDonald's functions (see Eq.
(\ref{diK})).}
\end{minipage}
\vspace{10mm}
\end{figure}
\begin{figure}
\begin{minipage}[t]{8cm}
\centerline{\epsfig{file=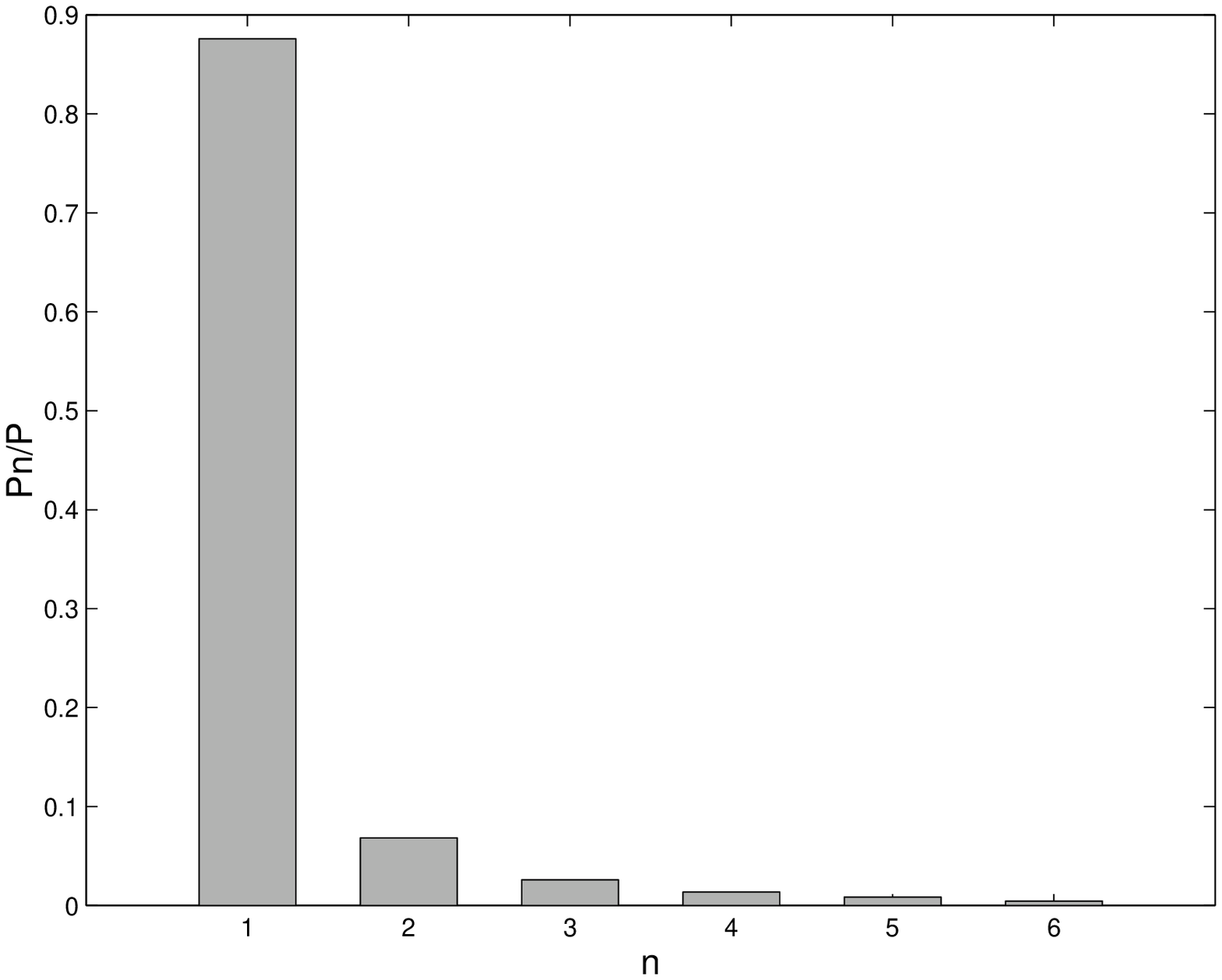,width=7cm}} \vspace{2.5mm}
\caption{Distribution of the dilaton radiation over the harmonic
number $n$.}
\end{minipage}\hspace{1cm}
\begin{minipage}[t]{8cm}
\centerline{\epsfig{file=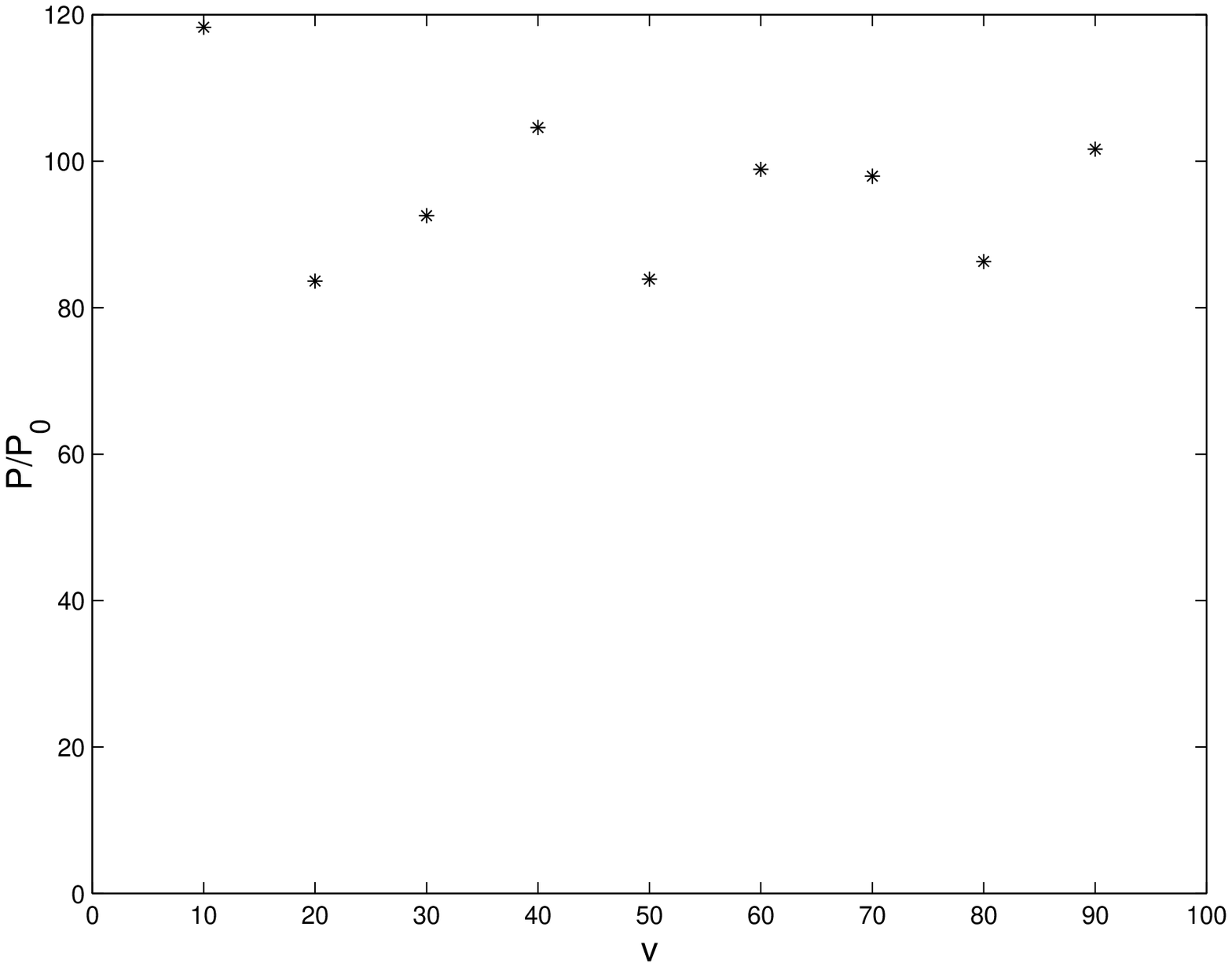,width=7cm}} \vspace{2.5mm}
\caption{Total radiation power as a function of the shape
parameter $v=\Omega R$.}
\end{minipage}
\vspace{10mm}
\end{figure}
\begin{figure}
\begin{minipage}[t]{8cm}
\centerline{\epsfig{file=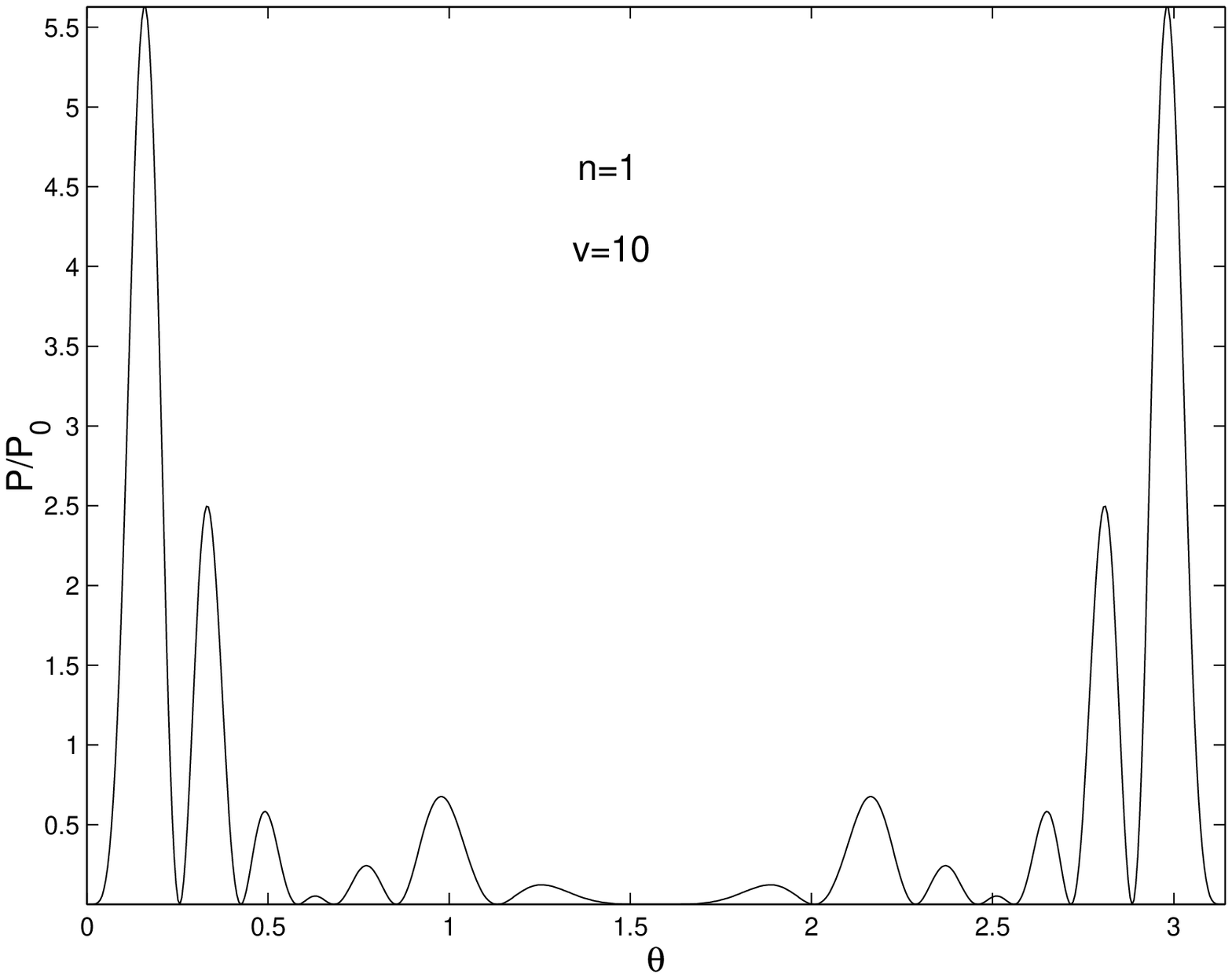,width=7cm}} \vspace{2.5mm}
\caption{Angular distribution of the first harmonic the gravitonal
radiation ($P_0=G \mu^2 R/ \Omega $)}.
\end{minipage}\hspace{1cm}
\begin{minipage}[t]{8cm}
\centerline{\epsfig{file=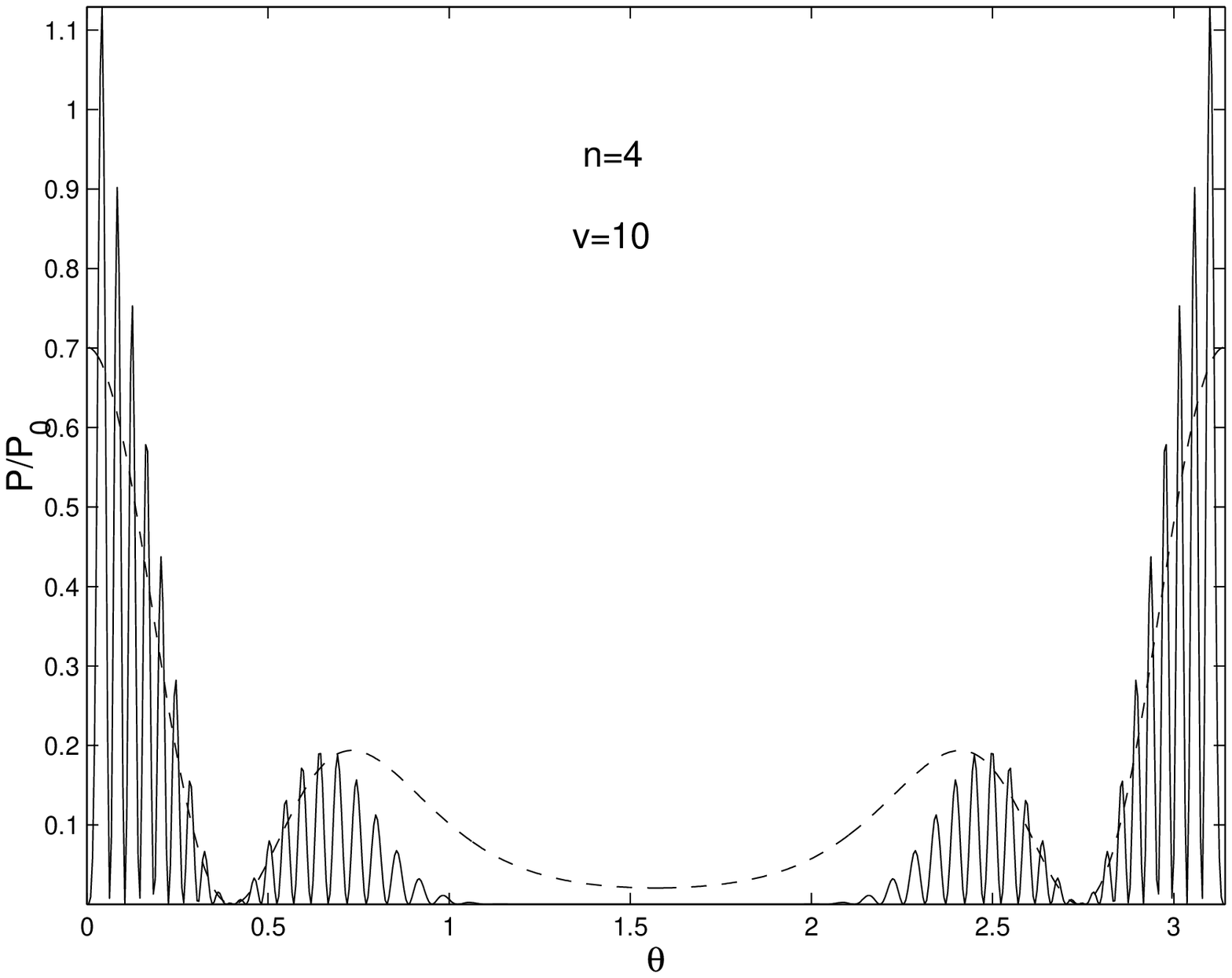,width=7cm}} \vspace{2.5mm}
\caption{Angular distribution of the 4-th harmonic the
gravitational radiation and its averaged approximation by
MacDonald's functions (see Eq. (\ref{grK}), $P_0=G \mu^2 R/ \Omega
$.}
\end{minipage}
\end{figure}
\begin{figure}
\begin{minipage}[t]{8cm}
\centerline{\epsfig{file=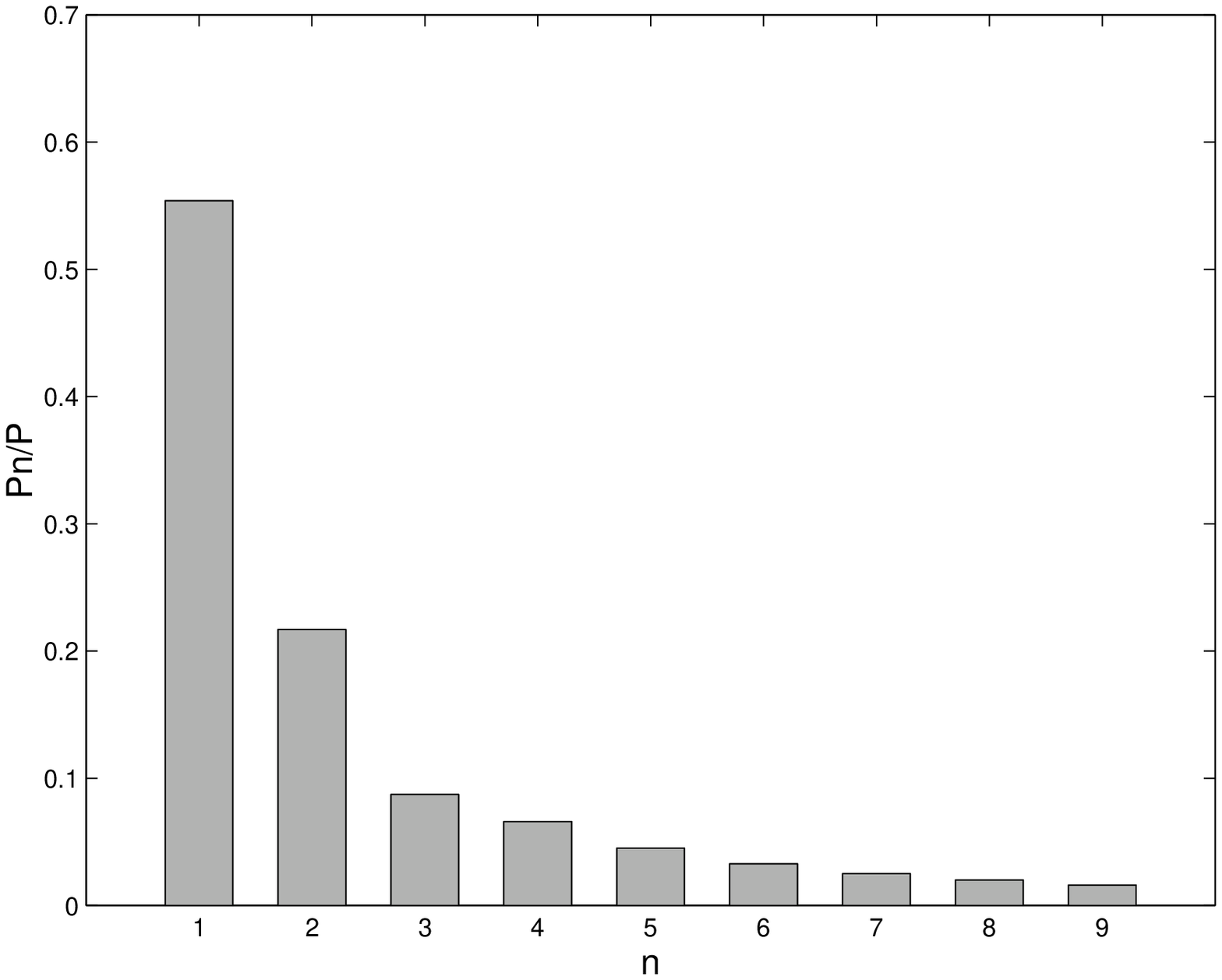,width=7cm}} \vspace{2.5mm}
\caption{Distribution of the gravitational radiation over the
harmonic number $n$.}
\end{minipage}\hspace{1cm}
\begin{minipage}[t]{8cm}
\centerline{\epsfig{file=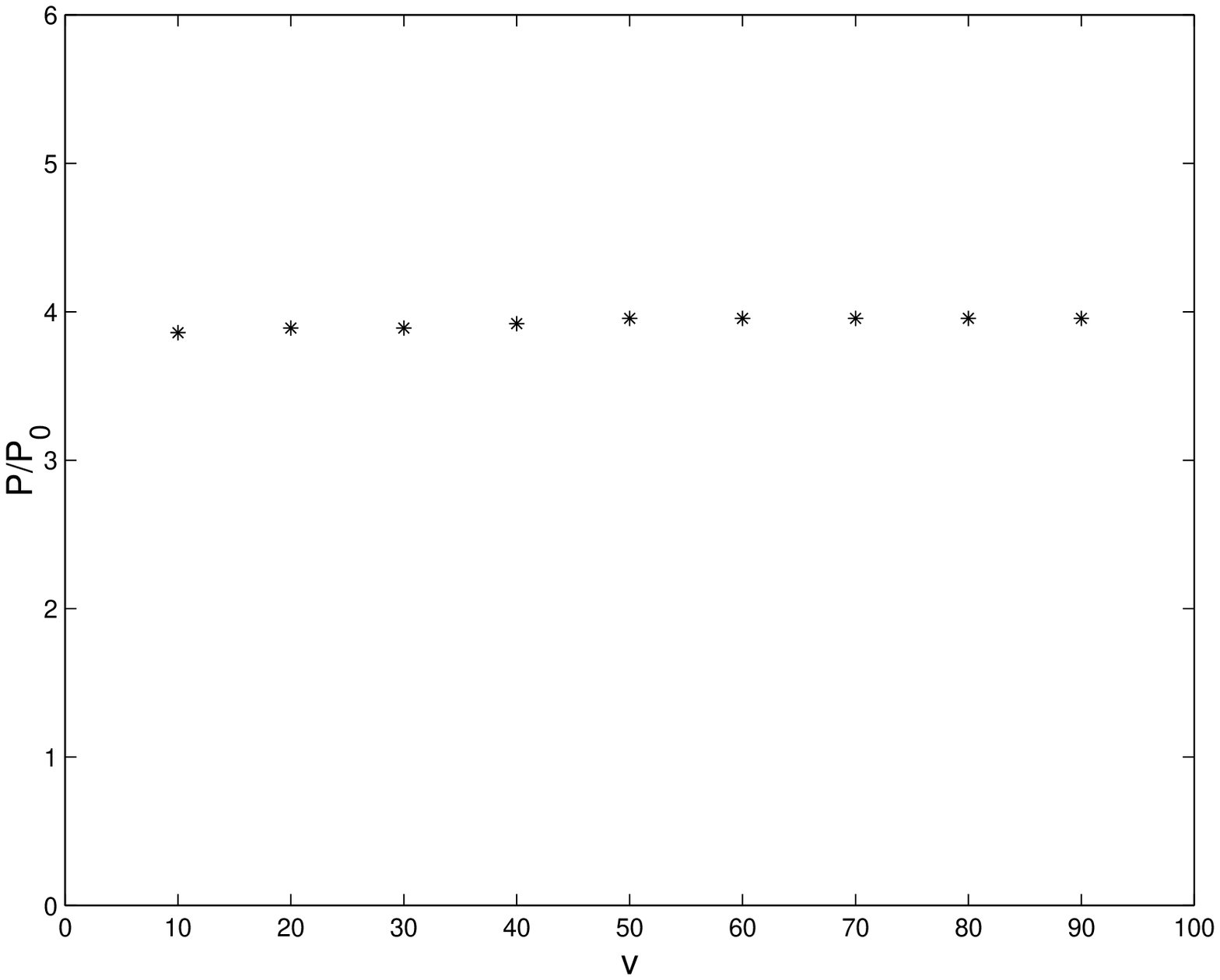,width=7cm}} \vspace{2.5mm}
\caption{Total gravitational radiation power as a function of the
shape parameter $v= \Omega R$.}
\end{minipage}
\end{figure}

\end{document}